\documentclass[10pt]{article}
\usepackage{physics}
\usepackage{graphicx}
\usepackage{times}
\usepackage{xcolor}
\usepackage[normalem]{ulem}
\usepackage{extarrows}
\usepackage{cite}
\usepackage[top=2cm,bottom = 2cm, left=1.3cm,right=1.3cm]{geometry}
\usepackage[resetlabels,labeled]{multibib}
\newcites{S}{References}
\newcommand{\biz}{\begin{itemize}}
\newcommand{\eiz}{\end{itemize}}

\newcommand{\be}{\begin{equation}}
\newcommand{\ee}{\end{equation}}
\newcommand{\bse}{\begin{subequations}}
\newcommand{\ese}{\end{subequations}}
\newcommand{\bpm}{\begin{pmatrix}}
\newcommand{\epm}{\end{pmatrix}}

\twocolumn

\title{A nondestructive Bell-state measurement on two distant atomic qubits}

\author{Stephan~Welte$^{1}$, Philip~Thomas$^{1}$, Lukas~Hartung$^{1}$, Severin~Daiss$^{1}$,\\ Stefan~Langenfeld$^{1}$, Olivier~Morin$^{1}$, Gerhard~Rempe$^{1}$, Emanuele~Distante$^{1,\ast}$
\\
\normalsize{$^{1}$Max-Planck-Institut f{\"u}r Quantenoptik,}
\normalsize{Hans-Kopfermann-Strasse 1, 85748 Garching, Germany}
\\
\normalsize{$^{\ast}$ To whom correspondence should be addressed. Email: emanuele.distante@mpq.mpg.de}}

\normalsize{\date{}} 

\begin{document}\sloppy
\maketitle
\noindent \textbf{One of the most fascinating aspects of quantum networks is their capability to distribute entanglement as a nonlocal communication resource \cite{Wehner2018}. In a first step, this requires network-ready devices that can generate and store entangled states \cite{Kimble2008}. Another crucial step, however, is to develop measurement techniques that allow for entanglement detection. Demonstrations for different platforms \cite{Michler1996,Riebe2004,Barrett2004,Chou2005,Moehring2007,Hofmann2012,Nolleke2013,Bernien2013,Delteil2016,Sisodia2017,Starek2018} suffer from being either not complete, or destructive, or local. Here we demonstrate a complete and nondestructive measurement scheme \cite{Barrett2005,Gupta2007,Ionicioiu2007} that always projects any initial state of two spatially separated network nodes onto a maximally entangled state. Each node consists of an atom trapped inside an optical resonator from which two photons are successively reflected. Polarisation measurements on the photons discriminate between the four maximally entangled states. Remarkably, such states are not destroyed by our measurement. In the future, our technique might serve to probe the decay of entanglement and to stabilise it against dephasing via repeated measurements \cite{Misra1976,Facchi2008}}.

Joint measurements that detect entangled states of multiple stationary qubits are a backbone for the development of quantum networks. A prominent example are Bell-state measurements (BSMs) that detect the maximally entangled Bell states (BSs) of two qubits \cite{Michler1996}. The BSMs enable fundamental protocols such as quantum teleportation and entanglement swapping \cite{Pirandola2015} for the purpose of quantum-information transfer and entanglement distribution over the network. A novel and fascinating scenario occurs for measurements that are able to detect an entangled BS without disturbing it. Such nondestructive BSMs \cite{Barrett2005,Gupta2007,Ionicioiu2007} project any state of the measured qubits onto the detected entangled state. This allows one to repeatedly measure a BS, thus opening up a route towards a new class of applications. Among others, an intriguing perspective is the possibility to protect the entangled state of two distant network qubits against environment-induced decoherence via the quantum Zeno effect \cite{Misra1976,Facchi2008}. To this end, however, the measurement must be efficient and the detection time should be faster than the coherence time of the entangled state. Here the entangled state coherence is assumed to decay more slowly than exponentially in time. 

An ideal BSM should primarily be complete: it should be able to distinguish between all four BSs of two qubits, defined as $\ket{\Phi^\pm} = \frac{1}{\sqrt{2}}\left(\ket{\uparrow_z \uparrow_z} \pm \ket{\downarrow_z \downarrow_z}\right)$ and $\ket{\Psi^\pm} = \frac{1}{\sqrt{2}}\left(\ket{\uparrow_z \downarrow_z } \pm \ket{ \downarrow_z \uparrow_z}\right)$, where we have used $\{\ket{\uparrow_z},\ket{\downarrow_z }\}$ as the qubit computational basis. However, developing a BSM for distant stationary qubits which is at the same time complete and nondestructive poses great experimental challenges. For example, complete BSMs can be realised by making two qubits interact via a quantum gate before measuring each qubit separately \cite{Riebe2004,Barrett2004}. Although challenging, this protocol can be applied to qubits located in distant network nodes, provided that a nonlocal quantum gate is available \cite{Chou2018, Wan2019, Daiss2021}. In both scenarios, however, such a scheme projects the qubits onto separable states and thereby destroys the entanglement while detecting it. Alternatively, a BSM can be implemented by entangling each qubit with one photon, interfering the two photons on a beamsplitter and detecting them with single-photon detectors \cite{Chou2005,Moehring2007,Hofmann2012,Nolleke2013,Bernien2013,Delteil2016}. This is particularly convenient for qubits residing in separate nodes as the photons can travel in optical fibres. However, this scheme is intrinsically not complete as only two out of the four BSs can be revealed \cite{Lutkenhaus1999}. An upgrade of it that realises a nondestructive BSM which is also complete would require a photon-photon quantum gate \cite{Hacker2016,Tiarks2019} which is however hard to implement experimentally.

\begin{figure*}[ht]
 \centering
 \includegraphics[width=0.6\textwidth]{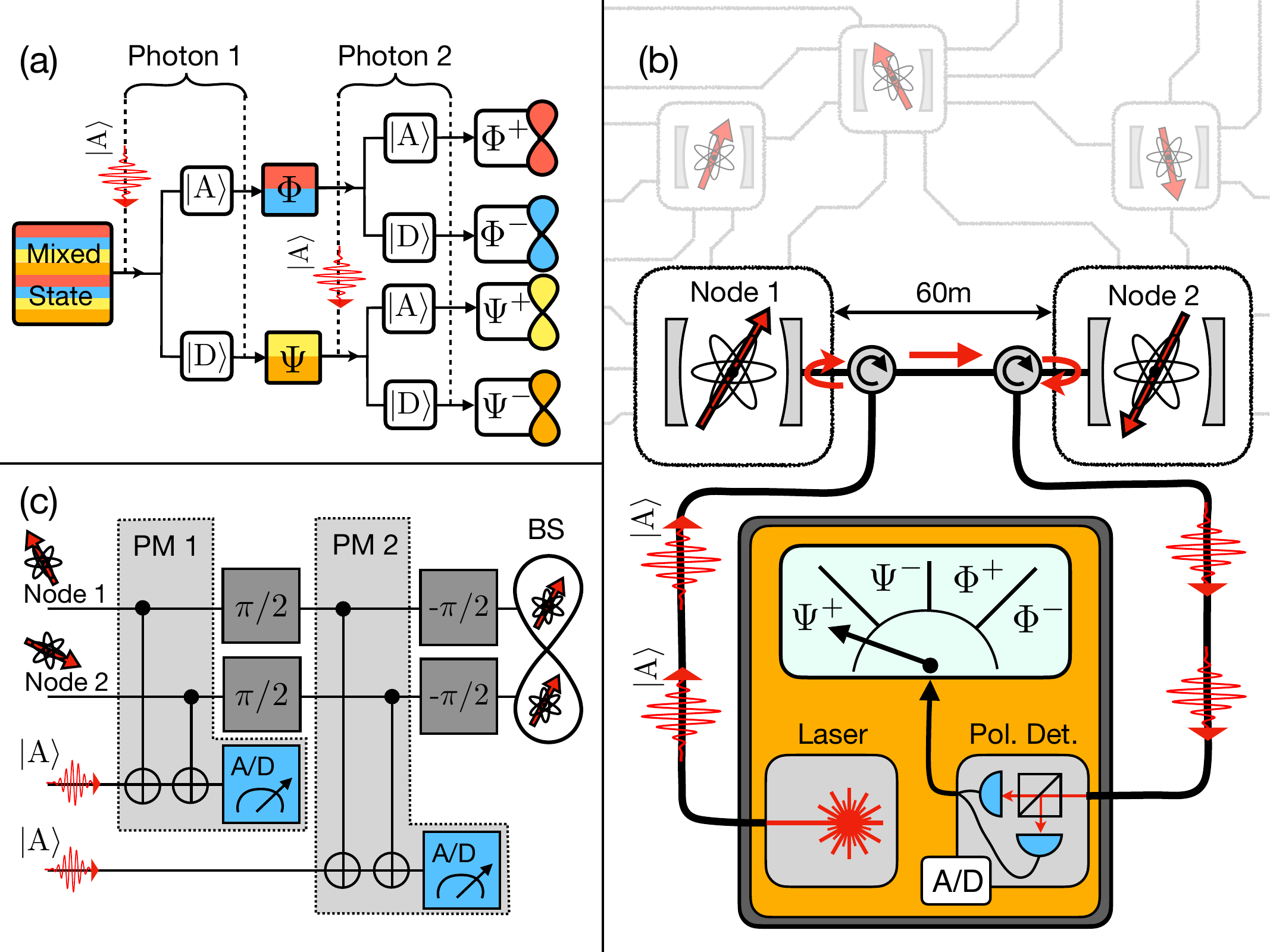}
 \caption{\textbf{Setup of the experiment} a) Decision tree to extract Bell states from a mixed state. Depending on the outcome of two photon-polarisation measurements (antidiagonal $\ket{A}$ or diagonal $\ket{D}$), the four Bell states result. b) Our setup comprises two single-atoms (red arrows) trapped at the centre of two resonators located at node 1 and node 2 and connected by an optical fibre. This forms a simple quantum network link that could be part of a larger quantum network architecture with additional nodes (greyed out). The measurement device for the nondestructive BSM, depicted in yellow, consists of a laser and a polarisation-sensitive set-up. It produces weak coherent pulses (red wiggly arrows) that are injected into the system, coupled to the resonators via optical circulators (depicted as circular arrows) and subsequently detected. c) Quantum circuit diagram of our protocol with two parity measurements (PM 1 and PM 2). Atom-photon gates are depicted as CNOT gates. The single-qubit rotations of the atoms are shown as grey boxes. The blue boxes represent the state detection of the two ancillary photons. The output Bell-state (BS) is represented by the 8-symbol.}
 \label{fig:fig1}
\end{figure*}

Here we demonstrate a different protocol that does not rely on photon interference or on a photon-photon gate but still realises a complete and nondestructive BSM of two atomic qubits located at two nodes of an elementary quantum network. We use two ancillary photons that travel between the nodes in an optical fibre link and interact sequentially with both atoms before being detected. Using suitable local qubit rotations, a single-photon state detection can distinguish between either the $\Phi$ and $\Psi$ or the $+$ and $-$ manifolds of the BSs (see Fig. \ref{fig:fig1}(a)). Two photons can then carry all the information to discriminate the four BSs. In essence, each photon implements a nondestructive parity measurement and two successive parity measurements together with the qubit rotations realise the nondestructive BSM scheme described in \cite{Gupta2007,Ionicioiu2007}. A similar protocol has been recently demonstrated on the IBM 5-qubit quantum processor chip \cite{Sisodia2017}. However, it relies on a stationary ancilla and is thus restricted to the measurement of qubits on the same chip. Conversely, our realisation employs travelling photons which enable the detection of entangled states of qubits located further apart. The only intrinsic limitation is the optical loss in the connecting fibre, which, at a suitable wavelength, can be small up to a few kilometres. Remarkably, the measurement time is much shorter than the coherence time of the atomic qubits. In a regime of reduced optical losses, this would allow to employ our scheme to stabilise the entangled state of distant qubits. Furthermore, as photons can connect multiple nodes, the presented scheme can be readily scaled up to generate and detect multi-qubit entangled states embedded in quantum networks \cite{Wang2013}.

Our experimental setup is shown in Fig. \ref{fig:fig1}(b). The qubits are two single $^{87}$Rb atoms located at node 1 and 2, respectively. They are physically separated by 21m and connected by a 60m single-mode optical fibre. Each atom is trapped at the centre of a high-finesse ($F=6\times 10^4$) optical cavity. Both cavities are single-sided such that light impinging on the input mirror will be effectively reflected back with high probability. The qubit space is formed by the two atomic ground states $\ket{\downarrow_z} = \ket{5S_{1/2},F=1,m_F = 1}$ and $\ket{\uparrow_z} = \ket{5S_{1/2},F=2,m_F = 2}$ and a pair of Raman lasers allows performing single-qubit rotations within this space. The cavities are actively stabilised to keep them resonant with the $\ket{\uparrow_z} \leftrightarrow \ket{e} = \ket{5P_{3/2},F=3,m_F = 3}$ atomic transition. The ancilla photons are polarisation qubits at $780\,\mathrm{nm}$ resonant with the $\ket{\uparrow_z} \leftrightarrow \ket{e}$ transition. To approximate single photons, we use strongly attenuated Gaussian laser pulses of 1 $\mathrm{\mu s}$ duration (FWHM) with an average photon number $\bar{n} \ll 1$. At each node, an optical circulator is used to couple light to the resonator and collect the reflected signal. Each photon interacts first with the atom-cavity system at node 1 before being collected and routed via the optical fibre to node 2. Here it interacts with the second atom-cavity system and it is finally guided to a photonic polarisation-detection setup realised with a combination of waveplates, a polarising beamsplitter and two superconducting nanowire single-photon detectors. 

The main building block of our nondestructive BSM is a CNOT gate between the atomic and the photonic qubit executed upon reflection of the photon from the resonator \cite{Duan2004,Reiserer2014}. The gate relies on a specific light shift of the atomic energy levels engineered such that only right circularly polarised photons $\ket{R}$ couple to the atom (via the $\ket{\uparrow_z} \leftrightarrow \ket{e}$ transition) while left circularly polarised photons $\ket{L}$ do not. Due to the strong atom-cavity coupling, an atom in the coupled state $\ket{\uparrow_z}$ prevents an $\ket{R}$ photon from entering the cavity and thus the photon is directly reflected back from the mirror. In contrast, if the atom is in the uncoupled state $\ket{\downarrow_z}$ or the photon is $\ket{L}$ polarised, the light circulates in the cavity before being reflected back. This results in a $\pi$ phase shift of $\ket{\uparrow_z, R}$  relative to the cases $\ket{\uparrow_z, L}$, $\ket{\downarrow_z, L}$,$\ket{\downarrow_z, R}$. Such atom-controlled $\pi$ phase shift realises a CNOT gate in the linear polarisation basis where a $\ket{\uparrow_z}$ atom flips an antidiagonal polarised $\ket{A} = 1/\sqrt{2}\left(i\ket{R}+\ket{L} \right) $ photon to its orthogonal diagonal polarisation $\ket{D} = 1/(\sqrt{2}i)\left(i\ket{R}-\ket{L} \right) $ and vice-versa, while the states $\ket{\downarrow_z, D}$ and $\ket{\downarrow_z, A}$ remain unchanged.

\begin{figure}[ht]
 \centering
 \includegraphics[width=0.4\textwidth]{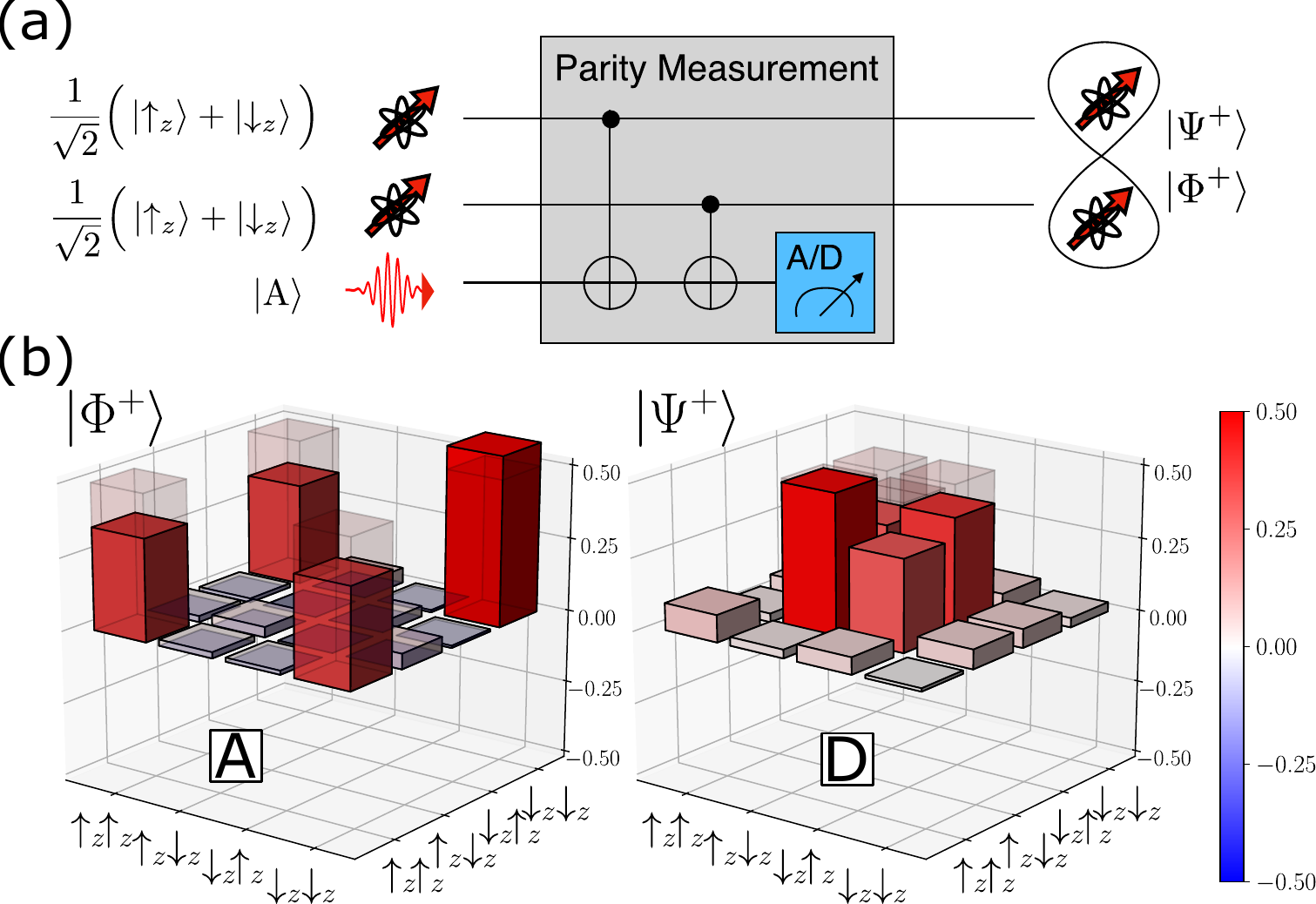}
 \caption{\textbf{Nondestructive parity measurement} (a) Quantum circuit diagram of the parity measurement. An antidiagonally polarised ancilla photon $\ket{A}$ reflects sequentially from node 1 and node 2 and implements a CNOT gate with each atomic qubit. The polarisation detection on the ancilla projects the atoms on a state with known parity. For the shown initial state, this results in one of the entangled states $\ket{\Phi^+}$ or $\ket{\Psi^+}$. (b) Real part of the two-atom density matrices corresponding to the two possible measurement outcomes ($\ket{A}$ or $\ket{D}$) of the photon polarisation. The atoms are initially prepared in the state $1/\sqrt{2}(\ket{\uparrow_z}+\ket{\downarrow_z})\otimes1/\sqrt{2}(\ket{\uparrow_z}+\ket{\downarrow_z})$, as indicated in part (a). The two density matrices show a large overlap with the entangled states $\ket{\Phi^+}$ and $\ket{\Psi^+}$ with fidelities $\mathcal{F}_A = (80.8 \pm 1.4) \%$ and $\mathcal{F}_D = (75.3 \pm 1.5) \%$, respectively (the errors indicate the standard deviation of the means).}
 \label{fig:carving}
\end{figure}

The quantum circuit diagram of our measurement scheme is shown in Fig. \ref{fig:fig1}(c). We employ antidiagonally polarised photons $\ket{A}$ that travel from node 1 to node 2. This results in two successive atom-photon CNOT gates, which, together with the final photon polarisation-detection, realise a nondestructive parity measurement on the atoms, as indicated by the greyed box in Fig. \ref{fig:fig1}(c) (for details, see Supplementary information, section 1). Specifically, a polarisation detection in $\ket{A}$ projects the atoms on an even parity state, a linear combination of the atomic product states $\ket{\downarrow_z \downarrow_z}$ and $\ket{\uparrow_z \uparrow_z}$ that preserve the photon's polarisation (here $\ket{x y}$ indicates state $\ket{x}$ and $\ket{y}$ on the first and second node respectively). Conversely, a polarisation detection in $\ket{D}$ projects the atoms onto an odd parity state, a linear combination of $\ket{\uparrow_z \downarrow_z}$ and $\ket{\downarrow_z \uparrow_z}$. As the BSs $\ket{\Phi^\pm}$ and $\ket{\Psi^\pm}$ have opposite parity, a single ancilla is sufficient to distinguish between them. To discern between all four states, we apply two local $\pi/2$ rotations to the atomic qubits after the first ancilla is detected. This effectively rotates $\ket{\Phi^-}$ into $\ket{\Psi^+}$ and vice-versa while leaving $\ket{\Phi^+}$ and $\ket{\Psi^-}$ unchanged (see Supplementary information, section 2). At this point, a second $\ket{A}$ polarised photon is employed to realise a second nondestructive parity measurement. There are four possible outcomes of the two combined photon polarisation-detections: $\ket{A,A}$, $\ket{A,D}$, $\ket{D,A}$ and $\ket{D,D}$ where the first (second) state in the ket indicates the detection result of the first (second) photon. These outcomes unambiguously identify $\ket{\Phi^+},\ket{\Psi^+}, \ket{\Phi^-}$ and $\ket{\Psi^-}$, respectively. Importantly, the atomic qubits are projected onto a known and still available entangled state, as only the two ancillary photons have been measured. At the end of the protocol, the two $\pi/2$ pulses are reversed with two additional $-\pi/2$ rotations on each atomic qubit.

We first show that detecting the polarisation of a single ancilla photon measures nondestructively the parity of the atomic state. To this end, we employ a coherent photon pulse with $\bar{n} = 0.1$ and prepare the atomic qubits in a specific initial state $\ket{\phi} = 1/\sqrt{2}(\ket{\uparrow_z}+\ket{\downarrow_z})\otimes1/\sqrt{2}(\ket{\uparrow_z}+\ket{\downarrow_z})$, as shown in Fig. \ref{fig:carving}(a). Since $\ket{\phi} =1/\sqrt{2}\left(\ket{\Psi^{+}} + \ket{\Phi^{+}}\right)$, a nondestructive parity measurement should always project the atomic qubits on an entangled state, either $\ket{\Phi^{+}}$ or $\ket{\Psi^{+}}$ depending whether $\ket{A}$ or $\ket{D}$ has been detected. We verify this by performing a full state-tomography on the two atomic qubits conditioned on a specific polarisation detection. Our results are presented in Fig. \ref{fig:carving}(b) where we show the reconstructed density matrices of the two possible final states. The measured fidelities with the ideal Bell states are $\mathcal{F}_A(\Phi^+) = (80.8 \pm 1.4) \%$ and $\mathcal{F}_D(\Psi^+) = (75.3 \pm 1.5) \%$ for a polarisation detection in $\ket{A}$ and $\ket{D}$, respectively. Here and in the rest of the paper, we define the fidelity $\mathcal{F}(x)$ of state $\rho$ with a pure state $\ket{x}$ as $\mathcal{F}(x)=\bra{x}\rho\ket{x}$ and the errors indicate the standard deviation of the means. The experimental limitations in this experiment are discussed further below. The measurement demonstrates that one photon polarisation-detection distinguishes between even and odd parity states and thus between the $\Psi$ and $\Phi$ manifolds of the BSs. 

\begin{figure*}[h]
 \centering
 \includegraphics[width=0.8\textwidth]{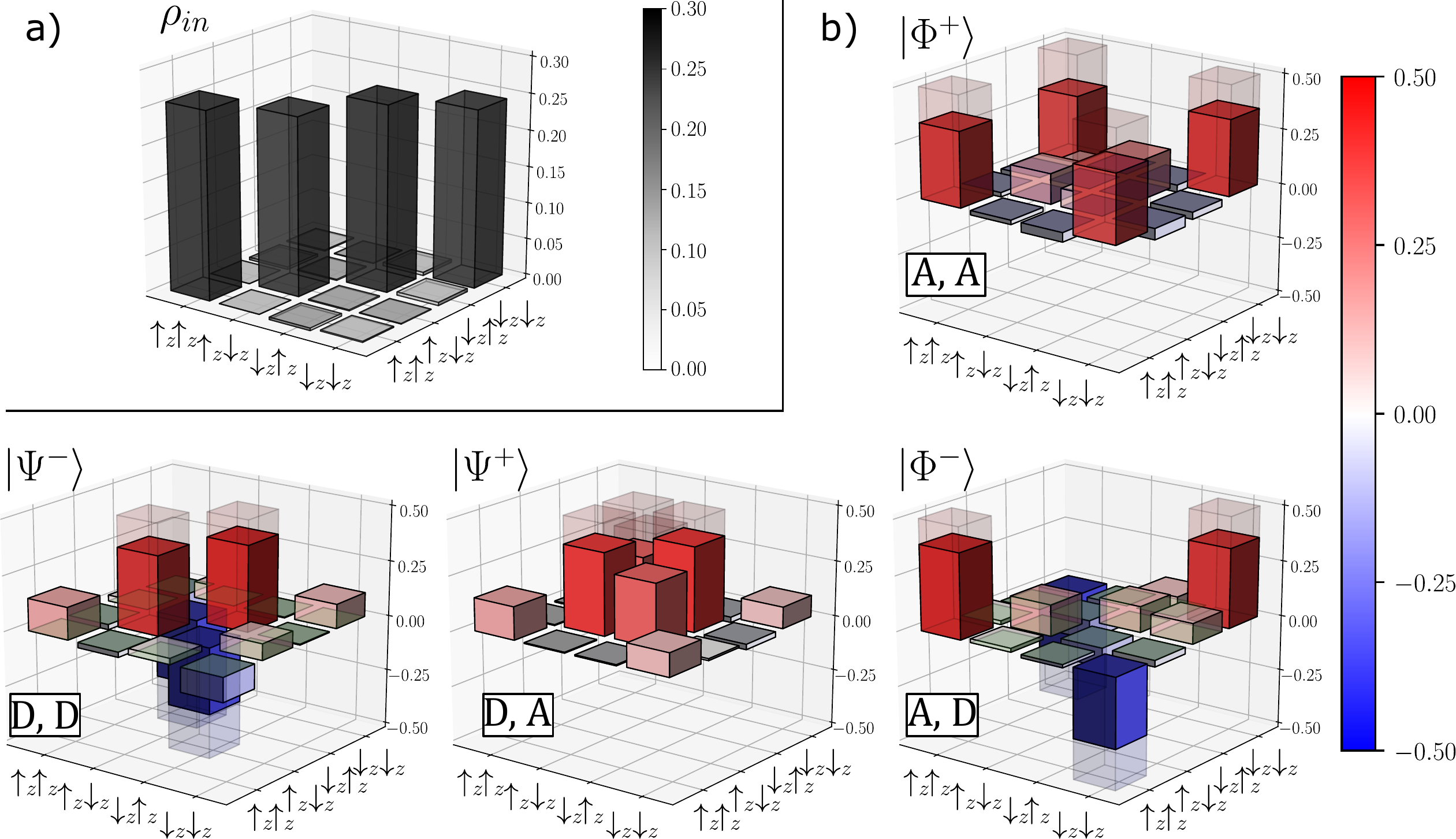}
 \caption{\textbf{Deterministic generation of entanglement from a mixed state} (a) Reconstructed density matrix of the fully mixed input state. (b) Real part of the reconstructed density matrices corresponding to the four possible measurement outcomes $\ket{A,A}$, $\ket{A,D}$, $\ket{D,A}$ and $\ket{D,D}$ of the two ancilla photons. Each density matrix shows a large overlap with one of the four BS with fidelities $\mathcal{F}_{A,A}(\Phi^+) = (65.3\pm 2)\%$, $\mathcal{F}_{A,D}(\Phi^-) = (68.8\pm 2)\%$, $\mathcal{F}_{D,A}(\Psi^+)=(66.4\pm 2)\%$ and $\mathcal{F}_{D,D}(\Psi^-)=(67.3\pm 2)\%$. Here the errors indicate the standard deviation of the means. The transparent bars indicate the density matrix of the ideal BSs.}
 \label{fig:output}
\end{figure*}

As a next step, we show the full protocol of the nondestructive BSM. We start by demonstrating a specific feature of such a measurement, which is that any state of the two atoms is projected onto the detected BSs and thus that our BSM always generates entanglement. To this end, we prepare the system in the highly mixed state shown in Fig. \ref{fig:output}(a). This is produced by randomly preparing each atom in one of the 6 initial states $\ket{\uparrow_z}$, $\ket{\downarrow_z}$, $\ket{\uparrow_x}=\frac{1}{\sqrt{2}}\left(\ket{\uparrow_z}+\ket{\downarrow_z}\right)$, $\ket{\downarrow_x}=\frac{1}{\sqrt{2}}\left(\ket{\uparrow_z}-\ket{\downarrow_z}\right)$, $\ket{\uparrow_y}=\frac{1}{\sqrt{2}}\left(\ket{\uparrow_z}+i\ket{\downarrow_z}\right)$ and $\ket{\downarrow_y}=\frac{1}{\sqrt{2}}\left(\ket{\uparrow_z}-i\ket{\downarrow_z}\right)$ where each state has equal probability to occur. We now perform the nondestructive BSM with two ancilla photon pulses that are both $\ket{A}$ polarised. We use an average photon number of $\bar{n} = 0.34$ for each pulse where the choice of a higher $\bar{n}$ allows to increase the success probability of the final photon detections. 
Conditioned on the four possible polarisation-measurement outcomes $\ket{A,A}$, $\ket{A,D}$, $\ket{D,A}$ and $\ket{D,D}$, we perform a full state-tomography on the two stationary qubits. The real part of the reconstructed density matrices are reported in Fig. \ref{fig:output}(b) (imaginary part is shown in Supplementary information Fig. S2). For each detection outcome $\ket{i,j}$, we compute the fidelity $\mathcal{F}_{i,j}(x)$ of the atomic qubits state with the expected entangled state $x$. We found them to be $\mathcal{F}_{A,A}(\Phi^+) = (65.3\pm 2)\%$, $\mathcal{F}_{A,D}(\Phi^-) = (68.8\pm 2)\%$, $\mathcal{F}_{D,A}(\Psi^+)=(66.4\pm 2)\%$ and $\mathcal{F}_{D,D}(\Psi^-)=(67.3\pm 2)\%$. This yields an average fidelity $\Bar{\mathcal{F}} = (66.9 \pm 2 \%)$ substantially larger than the classical threshold of $50\%$ thus certifying the genuine generation of entanglement. Our data show not only that the atomic qubits are always projected onto an entangled state but also that each combination of polarisation detection events is correlated with a different BS. Consequently, the two polarisation measurements can be used to discriminate unambiguously between the four BSs. Remarkably, the time between the first photon ancilla and the detection of the second ancilla is $9 \, \mathrm{\mu s}$, substantially shorter than the measured atomic coherence time of about $400 \, \mathrm{\mu s}$ for each atom \cite{Daiss2021}.

\begin{figure*}[h]
 \centering
 \includegraphics[width=0.7\textwidth]{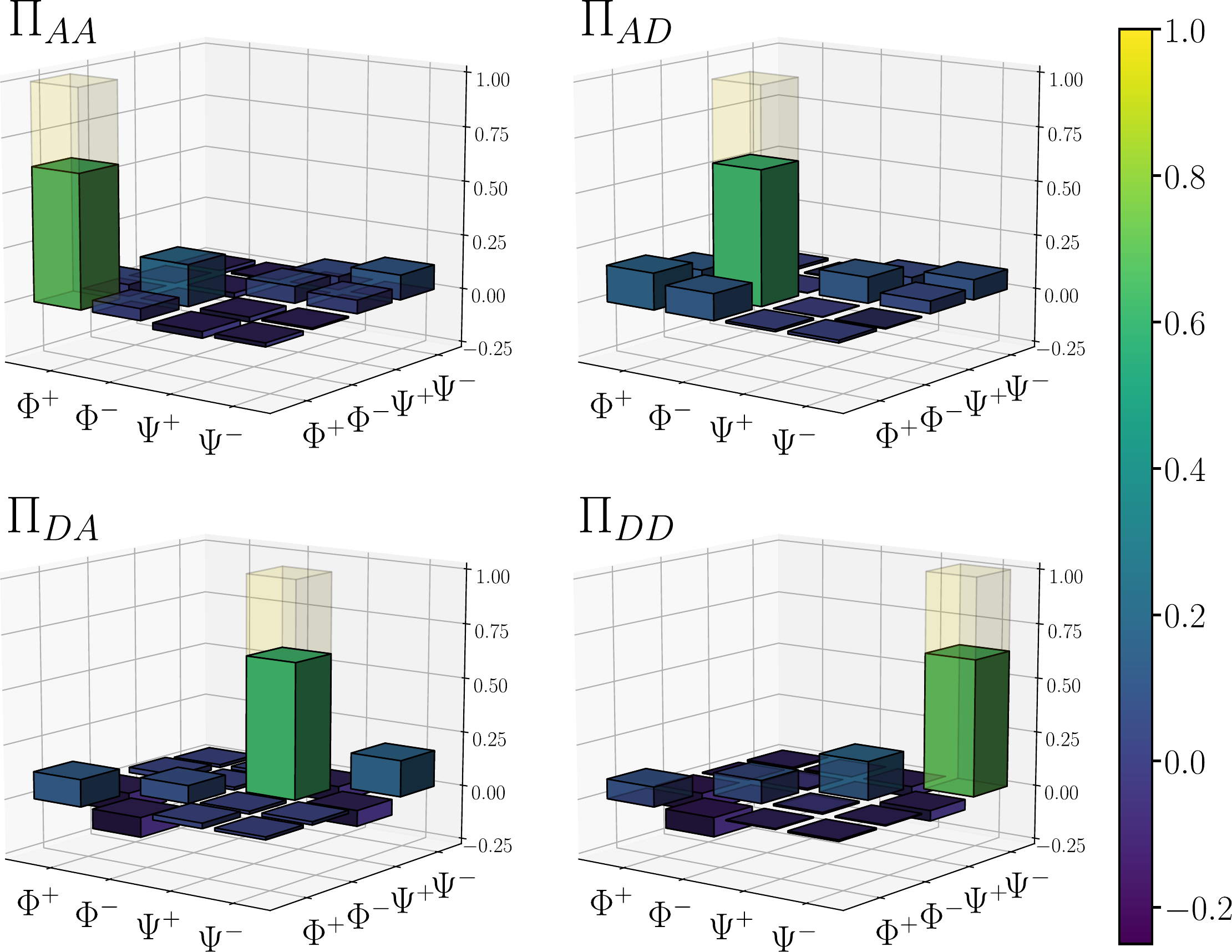}
 \caption{\textbf{Measurement tomography}. Real part of the reconstructed matrices describing the four operators $\{\Pi_j\}$ that constitute the POVM describing our measurement. The subscript $j\in\{AA, AD, DA, DD\}$ indicates the four possible outcomes of the polarisation detection. The operator matrices are represented in the Bell basis $\{\ket{\Phi^\pm}, \ket{\Psi^\pm}\}$. The transparent bars represent the density matrices for the ideal POVM of a nondestructive BSM.}
 \label{fig:fig4}
\end{figure*}

Finally, to unambiguously demonstrate the nondestructive character of our BSM, we perform a complete measurement tomography \cite{Lundeen2009}. Specifically, we reconstruct the four operators $\{\Pi_j\}$ which constitute the positive operator-valued measure (POVM) that fully describes our measurement. Here $j\in\{AA, AD, DA, DD\}$ is an index that labels the four possible outcomes of the two polarisation measurements. For an ideal BSM each $\Pi_j$ would correspond to a projector onto one of the four BSs, explicitly $\{\Pi_j\} = \{ \ketbra{\Phi^+}{\Phi^+}, \ketbra{\Phi^-}{\Phi^-}, \ketbra{\Psi^+}{\Psi^+}, \ketbra{\Psi^-}{\Psi^-} \}$. Following the quantum theory of measurement, for an initial density matrix $\rho$ of the two atomic qubits, the probability to detect an outcome $j$ is given by \cite{Lundeen2009}: 
\begin{equation}\label{POVM}
p_j = \mathrm{Tr}[\rho \Pi_j]. \end{equation}
To reconstruct $\{\Pi_j\}$, we prepare different combinations of initial states with each atomic qubit in one of the 6 states $\ket{\uparrow_z}$, $\ket{\downarrow_z}$, $\ket{\uparrow_x}$, $\ket{\downarrow_x}$, $\ket{\uparrow_y}$, $\ket{\downarrow_y}$. This results in a total of 36 possible initial states of the two atoms. For each of these states, we run our BSM protocol and measure the probabilities $p_j$. In this way we obtain the full information to invert the relation given by Eq. (\ref{POVM}) and to reconstruct the POVM $\{\Pi_j\}$. In Fig. \ref{fig:fig4}, we show the real parts of the four reconstructed operators $\Pi_j$ in the Bell basis (imaginary part is shown in Supplementary information Fig. S3). Our data show that each $\Pi_j$ has a large overlap with one of the four BSs. The measured fidelities are $\mathcal{F}_{DD}(\Phi^+) = (61.7 \pm 1.9) \%$, $\mathcal{F}_{DA}(\Phi^-) = (62.1 \pm 1.9)\%$, $\mathcal{F}_{AD}(\Psi^+) = (62.7 \pm 1.9)\%$ and $\mathcal{F}_{AA}(\Psi^-) = (62.7\pm 1.9) \%$ where we have used the notation $\mathcal{F}_j(x) = \bra{x}\Pi_j \ket{x}$. These data demonstrate that the polarisation measurement of the ancillary photons is effectively equivalent to a measurement projecting on the Bell basis of the atomic qubits. 

We numerically simulated our experiment including all known experimental imperfections (see Supplementary information section 3). From the simulation we deduce the impact of the different imperfections on the measured fidelities. We identify two major contributions. First, the multi-photon component of the weak coherent pulses limits the underlying atom-photon CNOT gates \cite{Reiserer2014}. This causes an overall fidelity loss of $4.3 \%$ and $9.8\%$ for the data in Fig. \ref{fig:carving} and Fig. \ref{fig:output}, respectively. Second, the limited overlap of the cavity transversal modes and the mode of the fibres used to couple the ancilla photons in and out at both nodes gives a reduction of $5.1 \%$. The remaining contributions include atomic state preparation and measurement, preservation of polarisation in the optical link, atomic state coherence and to the limited coherent coupling of the atom to the cavity field compared to the dissipative processes in the system (losses of the resonator mirrors and atomic scattering rates). They are discussed in more detail in the Supplementary information section 4. We attribute the difference in the fidelities reported in Fig. \ref{fig:fig4} with respect to Fig. \ref{fig:output} to larger errors related to the preparation of the different atomic input states.

In contrast to all previously implemented BSMs \cite{Michler1996,Riebe2004,Barrett2004,Chou2005,Moehring2007,Hofmann2012,Nolleke2013,Bernien2013,Delteil2016,Sisodia2017,Starek2018}, the complete, non-local and nondestructive nature of the presented scheme might allow using the quantum Zeno effect \cite{Misra1976,Facchi2008} as a means to suppress the decoherence of any BS of two distant qubits. Explicitly, if the coherence of the entangled state decays slower than exponentially in time, repeated applications of our nondestructive BSM would continuously project the qubits with high probability to their initial entangled state, effectively evading the detrimental effect of the environment by freezing the quantum-state evolution. To this end, however, the BSM must be efficient and faster than the characteristic decay time of the entanglement. In our current implementation the efficiency is limited to $0.1 \%$, as it is strongly affected by the overall optical losses ($8.6 \%$ transmission from before the resonator at node 1 to a detected photon click) and the large vacuum component ($72 \%$) of the employed weak coherent state. Future improved setups could strongly mitigate optical losses, including detector inefficiencies. This will enable using an optimised sequence to render the scheme highly efficient even with weak coherent states. In this case, the first pulse could be repeatedly sent until a successful photon detection occurs because near-zero optical losses assure that no photon has interacted with the qubits unless one was detected. The unitary $\pi/2$ atomic rotation would then follow after which a second series of pulses would be sent until another photon-detection event occurs. Assuming $\bar{n}=0.1$ per pulse and $4 \mathrm{\mu s}$ of atomic rotation (see Methods), a successful measurement could be carried out in 24 $\mathrm{\mu s}$ on average, considerably shorter than our atomic coherence time. This could be even pushed down to 3 $\mathrm{\mu s}$ using a deterministic single-photon source and faster atomic qubit rotations \cite{Hacker2016}.

A future implementation of our scheme could also improve on our currently reported fidelities as they do not suffer from any fundamental limitation. In fact, by using a single-photon source, optimising the cavity-to-fibre transversal mode matching and suppressing the polarisation errors, the current average entanglement fidelity could be boosted from $\bar{\mathcal{F} }=67.9 \%$ to $\bar{\mathcal{F}} \approx 90 \%$. Further improvements are then possible by better controlling the atomic qubit state and reducing the fluctuations of the resonator frequencies. 

Finally, a fascinating avenue is to extend the described nondestructive BSM to more nodes \cite{Kimble2008,Wehner2018}. This could be straightforwardly realised as our implementation makes use of travelling photons that can connect many distant qubits to implement a multi-qubit parity measurement. Together with single-qubit rotations, this would enable the generation and detection of multi-qubit entangled GHZ states \cite{Wang2013} in a larger quantum network. Finally, we stress that the role of the photonic ancilla and the atomic qubits are interchangeable. A slight modification to the current experiment can therefore be used to build a nondestructive BSM for photons. This could also be extended to generate and detect multi-photon GHZ states \cite{Bouwmeester1999}. 
\section*{Methods}

The two single-sided Fabry-Perot resonators at node 1 and 2 are each made of two mirrors separated by 0.5 mm with transmissions of $T_1= (3 \mathrm{ppm}, 92 \mathrm{ppm})$ and $T_2= (4 \mathrm{ppm}, 101 \mathrm{ppm})$, respectively. The relevant cavity-QED parameters are $g,\kappa,\gamma = 2\pi \times(7.6,2.5,3.0)$ MHz and $g,\kappa,\gamma = 2\pi \times(7.6,2.8,3.0)$ MHz for node 1 and 2 respectively. Here we have indicated with $g,2\kappa,2\gamma$ the atom-cavity coupling strength, the cavity linewidth, and the atomic decay rate from the state $\ket{e}$. Both nodes operate in the strong-coupling regime where $g > \kappa, \gamma $. 

At each node, we can prepare any atomic qubit state and detect it in any basis. This is achieved by using a pair of lasers which can drive coherent Rabi oscillations between the states $\ket{\downarrow_z}$ and $\ket{\uparrow_z}$ via a two-photon Raman process \cite{Hacker2016}. In the presented experiments, a $\pi/2$ rotation is performed within $4\mathrm{\mu s}$. Slight intensity fluctuations and the finite bandwidth of the Raman pulse limit the state preparation fidelity to $98.5 \%$. The atomic state is measured using light resonant with the $\ket{\uparrow_z}\leftrightarrow\ket{e}$ transition which allows to discriminate between the state $\ket{\uparrow_z}$ and $\ket{\downarrow_z}$ in $<\!5\,\mathrm{\mu s}$ with a fidelity $>99.8 \%$ by collecting the fluorescence light through the cavity mode. The average qubit detection fidelity in different bases is then $98.3\%$, which includes a preceding Raman pulse of given amplitude and phase. 

At node 1, light emerges from a single-mode fibre, is reflected on a low reflectivity ($1.5 \%$) beamsplitter employed as a circulator and impinges on the first cavity. The overlap between the fibre transversal mode and the cavity mode is measured to be $92\%$. Upon reflection, light passes through an acousto-optical modulator which serves as a fast optical-path switch that separates atomic state-detection fluorescence-light from the spatial mode of the fibre connecting the two nodes. The latter is stabilised at regular time interval using piezoelectric fibre squeezers \cite{Rosenfeld2008} to avoid polarisation fluctuations between the nodes. It has a $95\%$ intrinsic transmission at the photon wavelength ($780\,\mathrm{nm}$), a $67\%$ transversal mode matching with the resonator at node 1 and it is connected to a fibre-based optical circulator which is used to couple the light to the cavity at node 2 and to collect it upon reflection. Including the circulator transmission (80 $\%$ for one passage) this leads to an overall $51 \%$ optical losses between the two nodes. The mode matching between the cavity at node 2 and the circulator is measured to be $98\%$. After being collected by the fibre-circulator, the light is outcoupled to free space and sent to a combination of waveplates and a polarising beamsplitter. At each port of the beamsplitter, two fibres collect the light and guide it to two superconducting nanowire single-photon detectors (detector efficiency $\eta \sim 90\%$ at $\lambda = 780\mathrm{nm}$). Including the second passage of the circulator, this leads to an overall detection efficiency of $50 \%$ with a dark-count rate of $9\mathrm{Hz}$.

At each node, a successful loading of an atom is heralded by imaging scattered cooling light with an objective (NA=0.4) on an EMCCD camera. When a single atom is present in each of the nodes, a digital signal triggers the experimental sequence which runs at 1$\mathrm{kHz}$ repetition rate. It starts with a $200\, \mathrm{\mu s}$ long optical pumping phase to prepare both atoms in the $\ket{\uparrow_z}$ state after which the atomic qubit states are initialised via a Raman pulse. The main protocol follows with the two ancillary Gaussian laser pulses interleaved by a $\pi/2$ rotation on both atoms and it ends with the atomic state detection. Between the Raman state initialisation pulse and the state detection the overall protocol lasts $36\, \mathrm{\mu s}$.

\bibliographystyle{naturemag}
\bibliography{references.bib}

\section*{Data Availability}
The data that support the findings of this study are available in Zenodo with the identifier "doi:10.5281/zenodo.4604775" (https://doi.org/10.5281/zenodo.4604775).

\section*{Acknowledgments}
We acknowledge fruitful discussions with Stephan D\"{u}rr and Lukas Knips. This work was supported by the Bundesministerium f\"{u}r Bildung und Forschung via the Verbund Q.Link.X (16KIS0870), by the Deutsche Forschungsgemeinschaft under Germany’s Excellence Strategy – EXC-2111 – 390814868, and by the European Union’s Horizon 2020 research and innovation programme via the project Quantum Internet Alliance (QIA, GA No. 820445). E.D. acknowledges support by the Cellex-ICFO-MPQ postdoctoral fellowship program.

\section*{Author Contributions Statement}
All authors contributed to the experiment, the analysis of the results, and the writing of the manuscript.

\section*{Competing Interests}
There is NO Competing Interest

\newpage
\section*{Supplementary Information}

\subsection*{Protocol for the nondestructive parity measurement}
The nondestructive parity measurement is implemented using the scheme shown in Fig. 2(a) of the main text. Each photon reflection implements a local atom-photon CNOT gate \citeS{Reiserer2014} described by the truth table
\begin{equation}\label{eq:cnot}
    \begin{aligned}
        \ket{\uparrow_z, A} &\rightarrow \ket{\uparrow_z, D}\\
        \ket{\uparrow_z, D} &\rightarrow \ket{\uparrow_z, A}\\
        \ket{\downarrow_z, A} &\rightarrow \ket{\downarrow_z, A}\\
        \ket{\downarrow_z, D} &\rightarrow \ket{\downarrow_z, D}.
    \end{aligned}
\end{equation}

Consider for simplicity a generic pure state of the two atomic qubits $\ket{\phi} = \alpha \ket{\uparrow_z \uparrow_z} +  \beta\ket{\downarrow_z \downarrow_z} + \gamma \ket{\uparrow_z \downarrow_z}+ \delta \ket{\downarrow_z \uparrow_z}$ together with a single-photon  ancilla initially in the state $\ket{A}$. Using eq. \ref{eq:cnot}, it is straightforward to prove that the global state of the two atomic qubits and the ancilla after the successive interaction with both nodes is
\begin{equation}\label{eq:parity_state}
\begin{aligned}
    \ket{\phi}\ket{A} \rightarrow &\left(\alpha \ket{\uparrow_z \uparrow_z} +  \beta\ket{\downarrow_z \downarrow_z}\right) \ket{A} + \\
    &\left(\gamma \ket{\uparrow_z \downarrow_z}+ \delta \ket{\downarrow_z \uparrow_z}\right)\ket{D}.
\end{aligned}
\end{equation}

This shows that even parity states preserve the initial polarisation of the ancilla, while odd parity states rotate it to an orthogonal state. The final polarisation detection thus projects the atomic qubit state onto a state with known parity. 

\subsection*{Protocol for the nondestructive Bell-state measurement}
Here we give a mathematical description of the full protocol to discriminate between the four BSs as implemented in our experiment and shown in Fig. 1(c) of the main text. We write the initial atomic qubit state using the Bell basis $\{\ket{\phi_j}\} = \{\ket{\Psi^\pm}, \ket{\Phi^\pm}\}$ with $j = {0,...,3}$

\begin{equation}\label{eq:state_init}
    \rho = \sum_{i,j} c_{ij} \ket{\phi_i}\bra{\phi_j}. 
\end{equation}

A first ancilla with the polarisation state $\ket{A}$ reflects successively from the two nodes and is then detected in the $\ket{A}$/$\ket{D}$ basis implementing a nondestructive parity measurement. Conditioned on the polarisation detection, the atomic density matrix can be computed using eq. \ref{eq:cnot} as 

\begin{equation}\label{eq:state_firstancilla}
    \begin{aligned}
    \rho_{D} =  &c_{00}\ket{\Psi^+}\bra{\Psi^+} + c_{01}\ket{\Psi^+}\bra{\Psi^-}+\\
                      &c_{10}\ket{\Psi^-}\bra{\Psi^+} + c_{11}\ket{\Psi^-}\bra{\Psi^-}\\ 
                     \\ 
    \rho_{A} =  &c_{22}\ket{\Phi^+}\bra{\Phi^+}+c_{23}\ket{\Phi^+}\bra{\Phi^-} +\\ 
                      &c_{32}\ket{\Phi^-}\bra{\Phi^+}+c_{33}\ket{\Phi^-}\bra{\Phi^-} 
    \end{aligned}
\end{equation}

where $\rho_p$ indicates the atomic qubits density matrix after a $p$ polarisation detection. The above equation shows that one ancilla discriminates between the Bell states $\ket{\Psi^\pm}$ and $\ket{\Phi^\pm}$. After this first photon, the protocol continues with a $\pi/2$ rotation of both atomic qubits. This performs the mapping
\begin{equation}\label{eq:global_pi2}
    \begin{aligned}
        R_{1,2}(\pi/2)  \ket{\Phi^+}& = \ket{\Phi^+}\\
        R_{1,2}(\pi/2) \ket{\Phi^-}&=\ket{\Psi^+}\\
        R_{1,2}(\pi/2) \ket{\Psi^+}&=\ket{\Phi^-}\\
        R_{1,2}(\pi/2) \ket{\Psi^-}&=\ket{\Psi^-}
    \end{aligned}
\end{equation}
where we have indicated with $R(\pi/2)_{1,2}$ the $\pi/2$ rotation on both nodes. The rotated atomic states $\rho^\prime_{D} =  R_{1,2}(\pi/2)\rho_{D}R^\dagger_{1,2}(\pi/2)$ and $\rho^\prime_{A}= R_{1,2}(\pi/2)\rho_{A}R^\dagger_{1,2}(\pi/2)$ can be expressed as
\begin{equation}\label{eq:state_firstancilla_rotated}
    \begin{aligned}
        \rho^\prime_{D}= & c_{00}\ket{\Phi^-}\bra{\Phi^-} + c_{01}\ket{\Phi^-}\bra{\Psi^-}+\\
                         & c_{10}\ket{\Psi^-}\bra{\Phi^-} + c_{11}\ket{\Psi^-}\bra{\Psi^-},\\
                         \\
         \rho^\prime_{A}= & c_{22}\ket{\Phi^+}\bra{\Phi^+}+c_{23}\ket{\Phi^+}\bra{\Psi^+} +\\
                         & c_{32}\ket{\Psi^+}\bra{\Phi^+}+c_{33}\ket{\Psi^+}\bra{\Psi^+}.
    \end{aligned}
\end{equation}
Now we implement a second parity measurement. Conditioned on the detected polarisation of the second ancilla photon, the state $\rho^\prime_{D}$ will be projected onto $\ket{\Phi^-}\bra{\Phi^-}$ or  $\ket{\Psi^-}\bra{\Psi^-}$, and the state $\rho^\prime_{A}$ onto $\ket{\Phi^+}\bra{\Phi^+}$ or $\ket{\Psi^+}\bra{\Psi^+}$ for an $\ket{A}$ or a $\ket{D}$ polarisation detection, respectively. To retrieve the original atomic state a final unitary rotation of $-\pi/2$ must be applied to both atoms. This inverts the mapping given in eq. (\ref{eq:global_pi2}) ensuring that any input Bell state remains invariant under the measurement process. 
\renewcommand{\thefigure}{S1}
\begin{figure*}[t]
  \centering
    \includegraphics[width=0.7\textwidth]{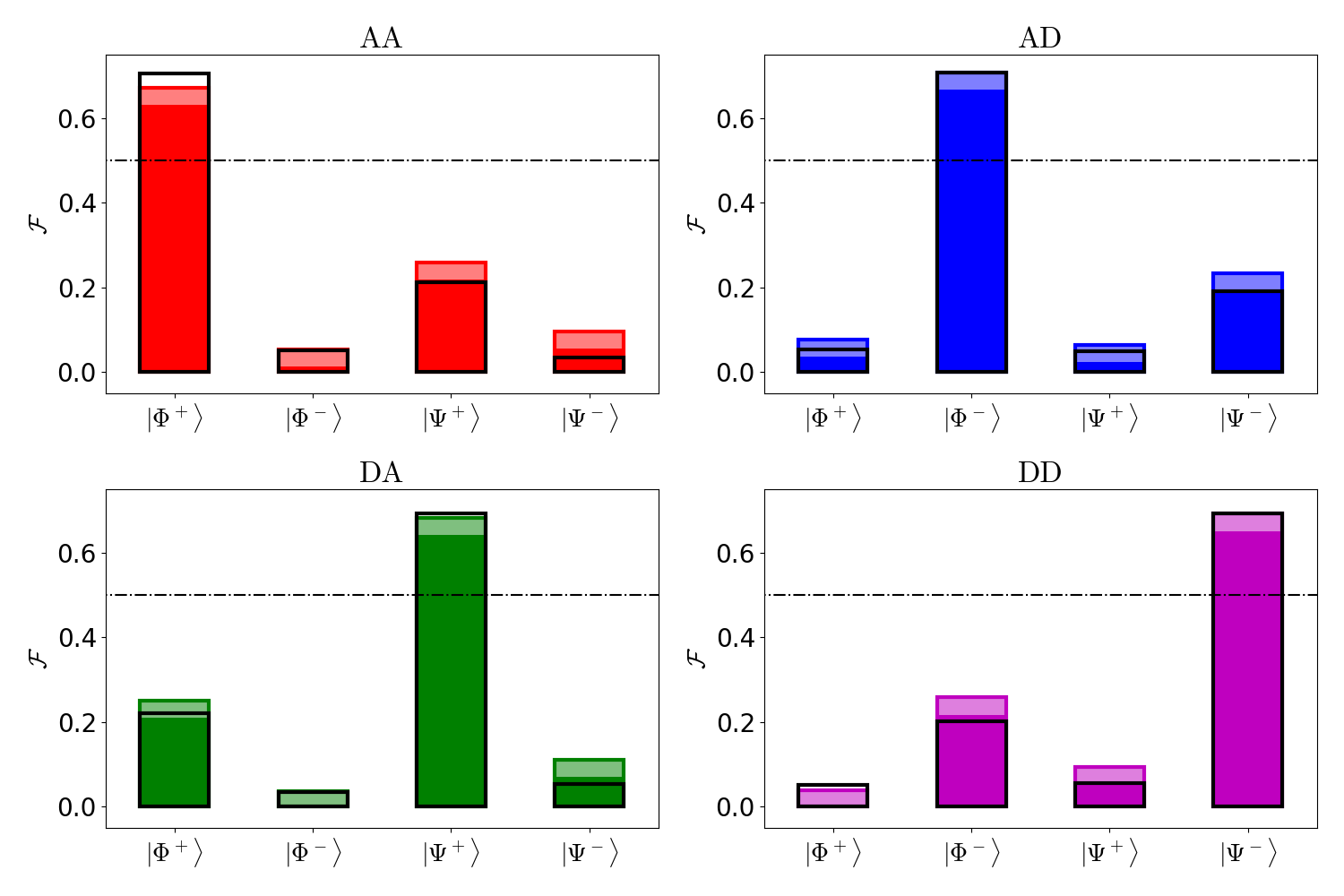}
    \caption{\textbf{Comparison of the simulated and measured fidelities}. Fidelity $\mathcal{F}$ of the simulated (solid black line) and measured (coloured bar) density matrices reported in Fig. 3(b) with the Bell state indicated on the horizontal axis. Each plot corresponds to a different outcome of the photon polarisation detection (indicated in the plot title). The light coloured area indicates the $\pm \sigma$ uncertainty of the measured fidelities.}
    \label{fig:output_comparison}
\end{figure*}

\renewcommand{\thefigure}{S2}
\begin{figure*}[htb]
  \centering
    \includegraphics[width=0.46\textwidth]{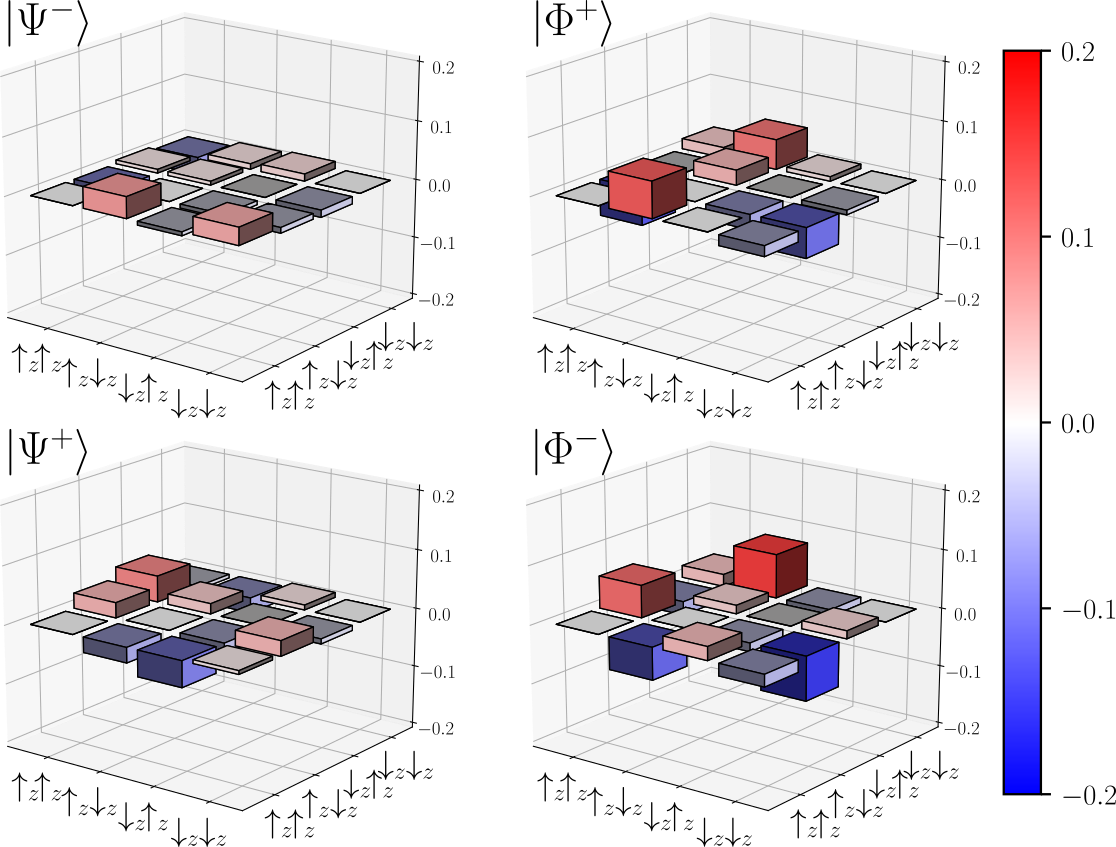}
    \caption{\textbf{Deterministic generation of entanglement from a mixed state}. Imaginary part of the reconstructed density matrices whose real part is shown in Fig 3 of the main text.}
    \label{fig:imaginary_Bell}
\end{figure*}

\renewcommand{\thefigure}{S3}
\begin{figure*}[htb]
  \centering
    \includegraphics[width=0.46\textwidth]{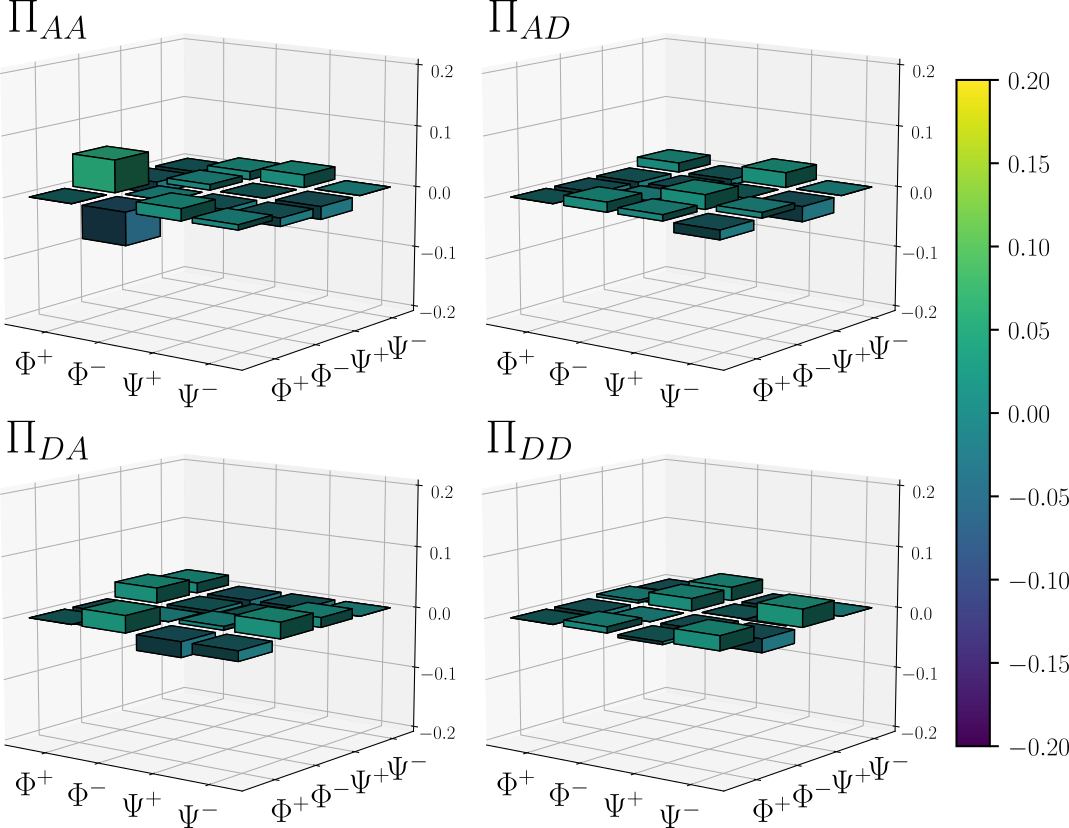}
    \caption{\textbf{Measurement tomography}. Imaginary part of the reconstructed matrices describing the four operators $\{\Pi_j\}$. The real part is shown in Fig 4 of the main text.}
    \label{fig:imaginary_POVM}
\end{figure*}

\subsection*{Numerical simulation}

 In this section, we give a more detailed description of the numerical simulation employed to estimate how the known experimental imperfections affect the measured fidelities. We base our simulation on the QuTip Python framework \citeS{Johansson2012}. At each node, we compute the interaction between a weak coherent pulse and a single atom trapped at the centre of a resonator using the input-output theory presented in\citeS{Hacker2019}. This allows us to include the measured cavity QED parameters, the limited transversal mode matching between the cavity modes and the optical fibres used to couple and collect light to the resonators (see Methods) as well as to account for the fluctuations of the cavity resonance frequency ($\pm 200$ kHz) due to the non-perfect cavity length stabilisation loop. We account for the finite transmission of the optical channel connecting the nodes using a beamsplitter and tracing out the unused mode. A polarisation beamsplitter in combination with a phase retarder is used to simulate polarisation dependent losses which arise mostly from the acousto-optical modulator employed as an optical switch. We include non-ideal single-photon detectors using an additional beamsplitter after node 2 that mixes the coherent state with uncorrelated thermal noise. The simulation takes into account the finite atomic state preparation and measurement fidelity as well as the finite fidelity of our Raman $\pi/2$ pulse used between the ancilla photons by additional depolarisation channels for the atomic qubit.
 
We first simulate the entangled state generation via the nondestructive parity measurement (data reported in Fig. 2(b)). Our simulation predicts a fidelity of $\mathcal{F}_A=78.7 \%$ and $\mathcal{F}_D=79.3 \%$ which compares well with the measured data. We then simulate the density matrices reported in Fig. 3(b) and we evaluate their fidelity with the four BSs. The comparison with the measured fidelities, shown in Fig. \ref{fig:output_comparison}, reveals that our simulation represents well the experimental data also in this second case. For completeness, in Fig. S2 we show the imaginary part of the measured density matrices (real part is shown Fig. 3(b)).

\subsection*{Analysis of experimental imperfections}

In addition to the multi-photon contributions of the employed weak coherent states and the transversal mode matching between the optical fibres and the resonators, we identified additional imperfections whose impact on the measured fidelities we discuss in this section. First, the limited atomic state preparation and measurement (SPAM) fidelity, which arises mostly from intensity fluctuations and the bandwidth of the Raman pulses (see Methods). This results in a reduction of fidelity of $4.1 \%$ for the data presented in Fig. 2. The data presented in Fig. 3(b), however, are insensitive to the state preparation fidelity, as the initial atomic state is a mixed state. Here SPAM errors gives $2.7\%$ of fidelity suppression. Still, in this latter case the Raman lasers are employed to perform the local atomic unitary rotations required for our protocol, accounting for an additional $2.6 \%$ fidelity reduction. The presented protocol relies on preserving the polarisation of the travelling ancilla photons. Therefore, we actively stabilise the 60 m fibre such that the polarisation of photons travelling in it is preserved with a fidelity $>99.9 \%$. However, additional optical elements between the two nodes cause polarisation-dependent losses. Together with polarisation fluctuations, this reduces the fidelity by $1.5\%$ and $3.5\%$ for Fig. 2 and Fig. 3(b), respectively. An additional contribution of $1.7\%$ (for data in Fig. 2) and $2.4\%$ (for data in Fig. 3(b)) arises from the finite coherence time of approximately $400 \mu s$ of the atomic qubits. As we employ magnetically sensitive $m_F$ states as atomic qubits, this is mostly caused by fluctuating external magnetic fields. Further contributions arise from fluctuations of the cavity resonance frequencies which, together with a slight birefringence of the cavities, account for a $2.6 \%$ and $3.0 \%$ fidelity reduction for data in Fig. 2 and Fig. 3(b), respectively. Finally, only a $0.5 \, \%$ ($0.8 \, \%$) reduction is given by our non-ideal cavity QED parameters for data shown in Fig. 2 (Fig. 3(b)).        

\subsection*{POVM tomography }

To reconstruct the positive operator-valued measure (POVM) that describes the BSM we follow a very similar strategy commonly used in quantum state tomography of a two-qubit state \citeS{Altepeter2004}. As stated in the main text, the POVM is composed of four operators $\{\Pi_j\}$ with $j\in\{DD, DA, AD, AA\}$. The probability of measuring an outcome $j$ conditioned on an input state $\rho_i$ of the atomic qubit is given by

\begin{equation}
    P(j\vert i)=\Tr(\rho_i\Pi_j).
    \label{eqn:povms}
\end{equation}

We probe the system with 36 known initial states $\rho_i$. These are all possible combinations of the states $\ket{\uparrow_x}$, $\ket{\downarrow_x}$,$\ket{\uparrow_y}$, $\ket{\downarrow_y}$, $\ket{\uparrow_z}$, and $\ket{\downarrow_z}$ for each of the atomic qubits. For each $\rho_i$, we record the probability of measuring an outcome $j$, resulting in a total of $36\times4$ probabilities. We then write the unknown two-qubit operator $\Pi_j$ using the Pauli matrices $\{\sigma_n\}$ with $n \in \{0,...,3\}$ (see \citeS{Altepeter2004})

\begin{equation}
\Pi_j=\frac{1}{2^2}\sum_{i,j=0}^{3}S^{(j)}_{n,m}\,\sigma_{n}\otimes \sigma_{m}
\label{eqn:povms_formula}
\end{equation}
where $S^{(j)}_{n,m}$ are the two-qubit Stokes parameters and $S_{0,0}=1$ due to normalisation. For each measurement outcome $j$, we can use eq. (\ref{eqn:povms_formula}) to invert eq. (\ref{eqn:povms}) and write each of the 15 unknown Stokes parameters as a function of the measured probabilities $P(j\vert i)$ similarly to \citeS{Altepeter2004}. In this way, we can reconstruct the associated positive-operator $\Pi_j$ for each outcome $j$ following eq. (\ref{eqn:povms_formula}). The real (imaginary) part of the reconstructed $\Pi_j$ is shown in Fig. 4 (Fig. S3).

\bibliographystyleS{naturemag}
\bibliographyS{referencesSupp.bib}
\end{document}